\documentstyle[12pt]{article}

\begin{document}
\begin{titlepage}
\begin{centering}
{\Huge\bf Influence of frustration on a d=3 diluted
 antiferromagnet: $Fe_{x}Zn_{1-x}F_{2}$}\\
\vspace{.5cm}
{\Large E.P. Raposo$^\dagger$, M.D. Coutinho-Filho
and F.C. Montenegro}\\
\vspace{.4cm}
{\large Departamento de F\'{\i}sica, Universidade
Federal de Pernambuco\\
Cidade Universit\'aria, 50670-901, Recife, PE, Brazil
\vspace{0.2cm}\\
\hspace{4.0cm}
$\dagger $ ecpr@npd1.ufpe.br}
\end{centering}
\begin{abstract}
\noindent
The influence of a frustrated bond on the magnetic
 properties of a d=3 uniaxial (Ising) b.c.c. diluted
 antiferromagnet, with emphasis in the compound
$Fe_{x}Zn_{1-x}F_{2}$, is investigated by a local
 mean-field numerical simulation. In particular we find
that the initial drop of the saturation staggered
 magnetization ($M_{S}$) with concentration follows a
percolation-like phenomenon characterized by an
exponent $\beta_{p}$. For the frustrated samples, however,
this regime is followed by a second one identified
by a ``long tail" effect such that
$M_{S}$ is zero only at the percolation threshold. Our
 numerical data also confirms a spin-glass phase near
 this threshold.
\end{abstract}
 solids.\\\\
\vspace{2.0cm}
\end{titlepage}
\large
\newpage
\noindent
Diluted antiferromagnets (DAFs) have been largely used
 as testing ground for theoretical models of random magnets[1].
In spite of this, a complete understanding of the microscopic
mechanisms behind the
rich variety of phases found in these systems is
still far from being accomplished. The particular nature of the
exchange interactions, the external applied field, as well as
the spatial anisotropy give rise to distinct magnetic properties in
real systems. With respect to the above parameters random systems
with uniaxial anisotropy and short-range interactions can be
classified by Ising models belonging to different universality
classes. In three dimensions (d=3) the insulating DAF
compound $Fe_{x}Zn_{1-x}F_{2}$, with a centred tetragonal
lattice, is one of the most studied experimental realizations
of such models[1-4]. Due to a strong single-ion anisotropy, pure
$FeF_{2}$ is considered an archetypal Ising system. The partial random
substitution of $Fe^{+2}$ by $Zn^{+2}$ ions in this compound make
it to behave magnetically as a random-exchange Ising model
(REIM). The application of a uniform field parallel to its
easy direction causes a crossover from REIM to the random-field
Ising model (RFIM) critical behavior[1,5]. For highly diluted samples,
random fields of strong magnitudes make the AF long-range order
(LRO) unstable and nucleates a ``glassy" phase in the upper part
of the ($H,T$) phase diagram[6]. In the vicinity of the percolation
threeshold ($x_{p} \approx 0.24$) a true Ising spin glass (SG) phase
was found to occur below a freezing temperature in a sample of
$Fe_{0.25}Zn_{0.75}F_{2}$ [4]. For the appearence of a SG
behavior randomness and frustration are considered essential
ingredients . However, the effects
of very weak frustrated interactions, eventually present
in short-ranged random magnets, have been {\it a priori}
neglected as compared with
 the ones caused by stronger interactions[2].
The opposite limit, i.e., systems with weak disorder and strong
frustrated interactions are also of considerable interest.\\
In a previous paper[3], we presented results of a zero-field
 ($H = 0$) numerical simulation which showed the
dramatic role played by a weak frustrated interaction
 in the antiferromagnetic ($M_{S}$) and spin-glass
($Q$) order parameters
 of the $Fe_{x}Zn_{1-x}F_{2}$, leading the system to the
SG phase observed[4] in the strong-dilution
 regime ($x \approx 0.25$) of this compound. Here, we
cover the influence of a
 frustrated bond in a centred tetragonal lattice of a
DAF (e.g. , $Fe_{x}Zn_{1-x}F_{2}$) at $H = 0$, for an extended range
of frustration
strenghts in the entire interval of dilution ($0 \leq
x \leq 1.0 $). $M_{S}(x)$ is shown to be strongly dependent on the
magnitude of frustration present in the whole range of $x$.
In particular it follows a percolation-like phenomenon
characterized by an
exponent $\beta_{p}$. For frustrated samples a ``long tail" effect
is identified such that
$M_{S}$ is zero only at $x_{p}$. On the other hand, $Q(x)$ is almost
independent on the frustration strenght, except for $x$ close to
$x_{p}$ where it suffers an abrupt increase in magnitude even for
the case of a very weak frustration.\\
We use a microscopic Hamiltonian suitable to describe a
 randomly diluted Ising antiferromagnet, with short-range
 exchange interactions
 $J_{\ell }$, given by:
\begin{equation}
{\cal H} = \sum_{<i,\delta_{\ell}>}J_{\ell}{\cal E}_{i}{\cal E}_
{i+\delta_{\ell}}S_{i}S_{i+\delta_{\ell}} ,
\end{equation}
where $S_{i} = \pm2$ (as in the $Fe^{+2}$),
${\cal E}_{i} = 0,1$ and $\ell $
 is summed up over the three sets of nearest neighbors
 in the centered tetragonal lattice shown in Fig. 1. We
 take $N = 2\cdot 30^{3}$ sites and use periodic
 boundary conditions. The local mean-field (LMF) method
 consists in solving iteratively the self-consistent
equations involving the thermally averaged spins $m_{i}$
 obtained through a variational minimization of the
MF-free energy[7], which gives for ${\cal E}_{i} = 1$
\begin{equation}
m_{i} = <S_{i}>_{T} = 2\tanh [(1/k_{B}T)\sum_{\delta_
{\ell}}J_{\ell}{\cal E}_{i+\delta_{\ell}}m_{i+\delta_
{\ell}}]  .
\end{equation}
The usual virtual crystal mean field (VCMF) result is
obtained assuming $m_{i} = m$ for every magnetic site.
The LMF simulation starts by choosing a random initial
configuration $\{S_{i},{\cal E}_{i}\}$ in
 the high-$T$(paramagnetic) phase. The system is cooled
in steps $\Delta T = 1$K through (2) and heated back by
the same amount up to a temperature of interest,
following a convergence criterion:
\begin{equation}
\frac{\sum_{i}[(m_{i})_{n} - (m_{i})_{n-1}]^{2}}{\sum_{i}
[(m_{i})_{n}]^{2}} \leq 10^{-6},
\end{equation}
where $n$ represents the $n$-th iteration.
We measured the staggered magnetization
\begin{equation}
M_{S}(T,x) = (2/N)\sum_{i}{\cal E}_{i}m_{i} ,
\end{equation}
with $i$ summed over a given sublattice, and the
spin-glass Edwards-Anderson (EA) order-parameter
\begin{equation}
Q(T,x) = (1/N)\sum_{j}{\cal E}_{j}<S_{j}^{2}>_{T},
\end{equation}
with $j$ taken through the whole lattice.
$M_{S}$ and $Q$ are averaged over 50 independent
 samples. In the simulations of the $Fe_{x}Zn_{1-x}F_{2}$
 compound we considered the exchange constants
$J_{\ell }$, shown in Fig. 1, keeping unaltered the
 experimental
 ratios $j_{1} \equiv J_{1}/J_{2} = -0.013$ and
$j_{3} \equiv J_{3}/J_{2} = +0.053$ measured[8] for
$Fe_{x}Zn_{1-x}F_{2}$, such as to fix the N\'eel temperature of
the pure system, $T_{N}(x = 1) = 77.8$ K. This
procedure appears to be quite realistic for
application in $Fe_{x}Zn_{1-x}F_{2}$, as
 early pulse experiments[9] in this system support
no $x$-dependence of the total exchange interaction
 in the entire concentration range measured
 ($0.2 \leq x \leq 1.0$).\\ The $x$-dependence of
the normalized sublattice magnetization
$M_{S}(x)/M_{S}(x = 1)$ is shown in Fig. 2, varying
the frustrated interaction in the whole interval
$0 \leq j_{3} \leq 1$. Data were obtained at
 $T/T_{N}(x=1) = 0.18$ and $H = 0$. The full line
represents the expected result of the VCMF method
for the specific case of $FeF_{2}$, where
$j_{3} = 0.053$ . As well known, the VCMF method
does not predict the existence of a percolation
threshold. Moreover, in this method
 the presence of a weak frustrated interaction
($j_{3} = 0.053$) plays no role in the linear dependence
 of $M_{S}$ with $x$. By way of contrast, in the LMF
 method we observe that even a very weak frustration
 ($j_{3} < 0.053$) reduces dramatically the values
of $M_{S}$ in the
 strong dilution regime($x < 0.5$), as is the case
 of $Fe_{x}Zn_{1-x}F_{2}$ [3]. However, as can be
 seen from the $j_{3} = 0.053$ curve, a weak frustration
 plays no effect on the AF order parameter $M_{S}$ of
 this compound for weak dilution ($0.6 < x < 1$). On
the other hand, a strong frustration ($j_{3} \approx 1$)
 may change the nature of the long range
 order even in the pure system, $x = 1$. $M_{S}(x)$
vanishes at the percolation concentration
$x_{p} = 0.247 \pm 0.010$, determined by the projection
 to the $x$-axis of the best fit curve from the $j_{3} = 0$
 data. This is in very good agreement with the value
$x_{p} = 0.243 \pm 0.010$, obtained[10] by series
 methods for non-frustrated b.c.c. lattices.
 In the LMF method $M_{S}$ becomes smaller
than the VCMF values for $x$ below a frustration-dependent
 value $x(j_{3})$. We perceive from the data of Fig. 2
that $x(j_{3})$ increases for increasing $j_{3}$, but
reaches the saturation value $x = 1.0$ for $j_{3} < 1.0$
. For non-zero frustration strenghts the $M_{S}$ curves
 display tails toward the percolation threshold, after
sudden initial drops starting at $x(j_{3})$. Close to
 the percolation threshold,
 the antiferromagnetic order parameter have the asymptotic
 form $M_{S} \sim (x - x_{p})^{\beta_{p}}$, where
$\beta_{p}$ is the percolation
 exponent associated to the order parameter. The best
 fitting of this law, with $j_{3} = 0$, gives the
value $\beta_{p} = 0.63 \pm 0.11$ . We notice the same
behavior in the frustrated systems
for which we find $\beta_{p} = 0.59 \pm 0.10$ for $j_{3} = 0.053$
 ($FeF_{2}$), and $\beta_{p} = 0.62 \pm 0.09$ for
 $j_{3} = 0.265$.
Series expansion[11] and
renormalization group[12] methods yield values
for $\beta_{p}$ between 0.4 to 0.5 . One should
 stress, however, that $M_{S}$ vanishes only at
 $x = x_{p}$ for the whole interval
$0 < j_{3} < 1.0$ .\\
The normalized Edwards-Anderson order parameter
 $Q(x)/Q(x=1)$ is plotted versus $x$, for several
values of $j_{3}$ in Fig. 3. The linear dependence
of $Q(x)$ with $x$, found for $x > 0.4$, is not disturbed
 even for the strongest magnitude of frustration used
 in this work ($j_{3} = 1.0$).
 However, close to the percolation threshold even an
evanescent frustration, as is the case of $Fe_{x}Zn_{1-x}F_{2}$,
 causes a pronounced increasing
 in $Q(x)$, as shown in the inset of Fig. 3. The values
 $M_{S}(x=0.25) \approx 0$ and $Q(x=0.25) \not= 0$ found
 in this LMF simulations for $j_{3} = 0.053$ explains[3]
 the spin glass phase experimentally detected in
$Fe_{0.25}Zn_{0.75}F_{2}$.
 For $x < 0.24$ however, both $M_{S}$ and $Q$ values are
 negligibly small (see Figs. 2 and 3, respectively) for
 the whole T interval measured in the present work
(2K $< T < 150$K). So, the LMF method does not support
 a true spin-glass phase ($Q \not= 0, M_{S} \approx 0$)
 for very low concentrations of magnetic ions, in
conformity with the short-range character of the exchange
 interactions.\\ In the LMF approach $T_{N}(x)$ is
determined by the inflection point of the $M_{S}(x)$
versus $T$ curve (see Ref. [3]). The $x$ dependence
 of the normalized N\'eel temperature $T_{N}(x)/T_{N}
(x=1)$ , obtained by VCMF and LMF methods, is compared
 with experimental results for the $Fe_{x}Zn_{1-x}F_{2}$
 compound in the phase diagram of Fig.4. The LMF
predictions for $T_{N}(x)$ are in good agreement with
 experimental data in the compound $Fe_{x}Zn_{1-x}F_{2}$
 for the whole range of $x$, in spite of the increasing
 error bars in the vicinity of the singular point
$x_{p}$. This large uncertainty is due to numerical
difficulties to obtain $M_{S}$ at very low temperatures
 ($T/T_{N}(x=1) < 0.17$) in the strong-dilution regime
 ($0.24 < x < 0.31$). Notice that $T_{N}(x)$ is
virtually unchanged even by the presence of moderate
 strenghts of frustration (see Fig. 2). This is explained
 by the fact that in spite of a sudden reduction on
 the magnitude of the antiferromagnetic
order parameter $M_{S}(x)$ when a frustrated interaction
 ($0 < j_{3} < 1.0$) is present, $M_{S}(x) \not= 0$
for all $x > x_{p}$ (see Fig. 2).
However, we cannot exclude the possibility that the
inclusion of thermal fluctuations may destroy such tiny
residual $M_{S}(x)$, leading $T_{N}(x)$ to vanish for
$x > x_{p}$, as experimentally observed in $Fe_{x}Zn_{1-x}F_{2}$
 [13].\\
Our reported results show that the presence of a small
frustrated interaction plays no role on the magnetic
 properties of a DAF system for weak dilution. However
 in the moderately and strongly diluted regimes it
causes major effects on a variety of magnetic properties,
 some of which have been identified in this work.\\
This work was supported by FINEP, CNPq, CAPES and FACEPE
 (Brazilian Agencies).
\newpage
{\bf References}\\
\begin{itemize}
\item{[1]} For a review of random
 disorder in magnetic systems, see: D.P. Belanger and A.P.
 Young, J. Magn. Magn. Mat. 100 (1991) 272. For the spin-glass problem,
 see e.g. : K.H. Fisher and J.A. Hertz,
 Spin Glasses (Cambridge University Press, London, 1991).

\item{[2]} V. Jaccarino and A.R. King, in: New Trends in
 Magnetism, ed. M.D. Coutinho-Filho and S.M. Rezende (World
 Scientific, Singapore, 1989) p. 70.

\item{[3]} E.P. Raposo, M.D. Coutinho-Filho and F.C.
Montenegro, Europhys. Lett. 29 (1995) 507.

\item{[4]} F.C. Montenegro, M.D. Coutinho-Filho and S.M.
 Rezende, J. Appl. Phys. 63 (1988) 3755; Europhys. Lett.
 8 (1989) 382; S.M. Rezende, F.C. Montenegro, M.D.
Coutinho-Filho, C.C. Becerra and A.J. Paduan-Filho, J.
Phys. (Paris) 49 (1989) C8 1267.

\item{[5]} One should notice that it has been proposed that frustration
in diluted magnetic systems can mimic a RFIM, even in
a zero external field: J.F. Fernandez, Europhys. Lett. 5 (1988) 129; C.L.
Henley, Phys. Rev. Lett. 62 (1989) 2056; C. Wengel, C.L. Henley and
A. Zippelius, {\it Preprint} cond-mat/9509056.

\item{[6]} F.C. Montenegro, U.A. Leitao, M.D. Coutinho-Filho
 and S.M. Rezende, J. Appl. Phys. 67 (1990) 5243;
 F.C. Montenegro, A.R. King, V. Jaccarino, S-J. Han and
D.P. Belanger, Phys. Rev. B 44 (1991-I) 2155.

\item{[7]} G.S. Grest, C.M. Soukoulis and K. Levin, Phys.
 Rev. B 33 (1986) 7659.

\item{[8]} M.T. Hutchings, B.D. Rainford and H.J.
Guggenheim, J. Phys. C 3 (1970) 307.

\item{[9]} A.R. King, V. Jaccarino, T. Sakakibara, M.
 Motokawa and M. Date, Phys. Rev. Lett. 47 (1981) 117;
 J. Appl. Phys. 53 (1982) 1874.

\item{[10]} M.F. Sykes and J.W. Essam, Phys. Rev. 133
(1964) A310.

\item{[11]} H.E. Stanley, J. Phys. A 10 (1977) L211, and
 references therein.

\item{[12]} R.G. Priest and T.C. Lubensky, Phys. Rev.
 B 13 (1976) 4159 and 14 (1977) 5125 (erratum); D.J.
 Amit, J. Phys. A 9 (1976) 1441.

\item{[13]} I.B. Ferreira, A.R. King and V. Jaccarino,
 Phys.Rev. B 43 (1981) 10797; D.P. Belanger and H.
 Yoshizawa, Phys.Rev. B 47 (1993) 5051.

\end{itemize}
\newpage
\noindent

{\bf Figure Captions}\\\\
\begin{itemize}
\item[Fig.1] b.c.c. magnetic structure, illustrating the
first three nearest-neighbors exchange interactions
between magnetic ions.
\footnote{Unfortunately we don't have a ps file of this figure.
However, it is also reproduced in Ref.[3] .}

\item[Fig.2] $x$-dependence of normalized sublattice
 magnetization $M_{S}(x)$, at $T/T_{N}(x=1) = 0.18$ and
 $H = 0$, for various exchange ratios $j_{3} = J_{3}/J_{2}$,
 measured by LMF simulations (symbols) and VCMF method
 (full line).

\item[Fig.3] $x$-dependence of normalized spin-glass
 Edwards-Anderson order parameter $Q(x)$, at
 $T/T_{N}(x=1)=0.18$ and $H = 0$, for various exchange
 ratios $j_{3} = J_{3}/J_{2}$, measured by LMF simulations.

\item[Fig.4] $T_{N}(x)/T_{N}(1)$ vs. $x$ using VCMF
 (full line) and LMF simulations in the frustrated
 ($\bullet$) and non-frustrated ($\Box$) cases, $H = 0$.
 Experimental data ($\bigtriangleup$) of
$Fe_{x}Zn_{1-x}F_{2}$ are plotted for comparison
 (Ref. [13]). $x_{p} = 0.247 \pm 0.010$ indicates
the percolation threshold.

\end{itemize}
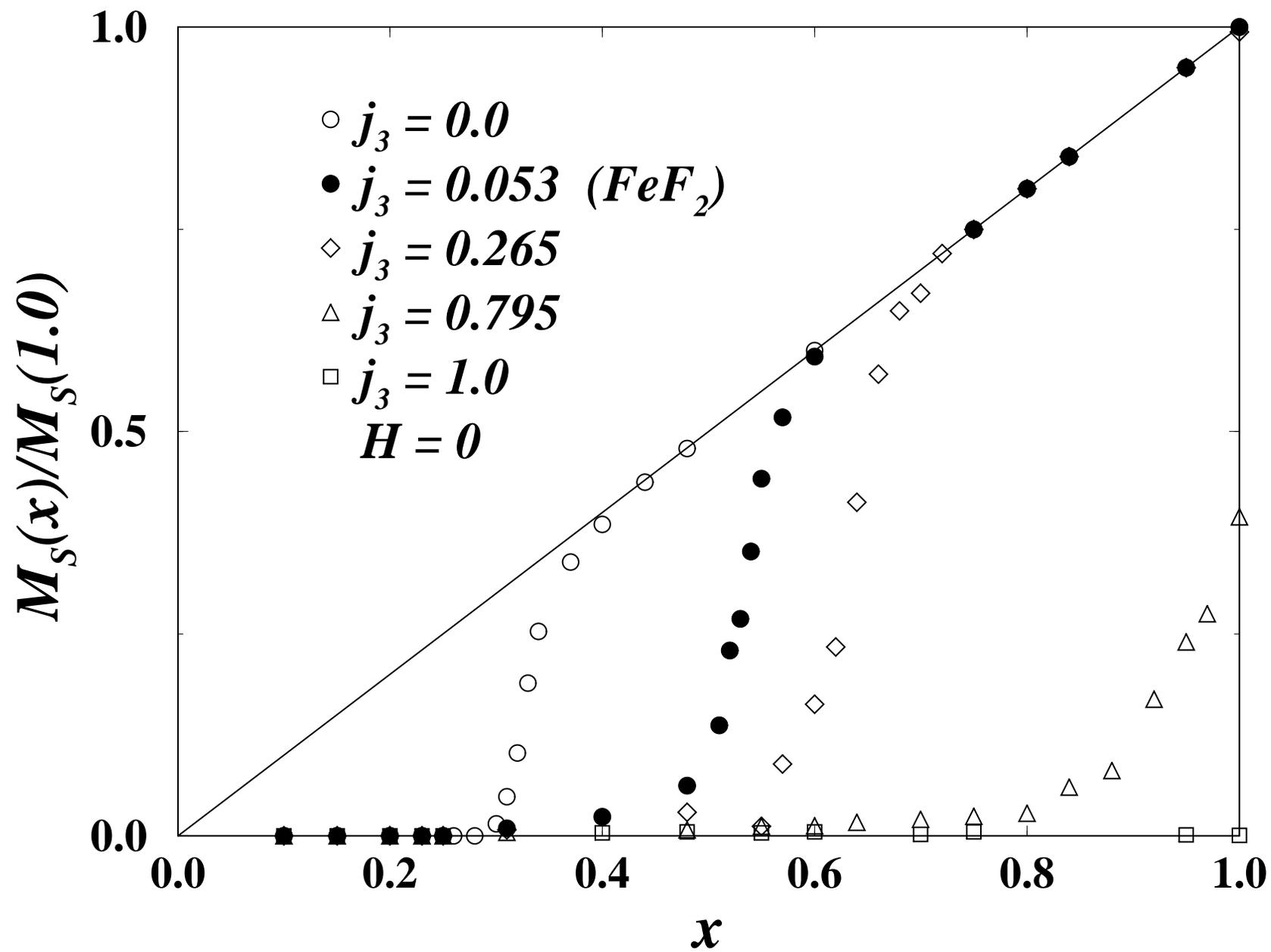
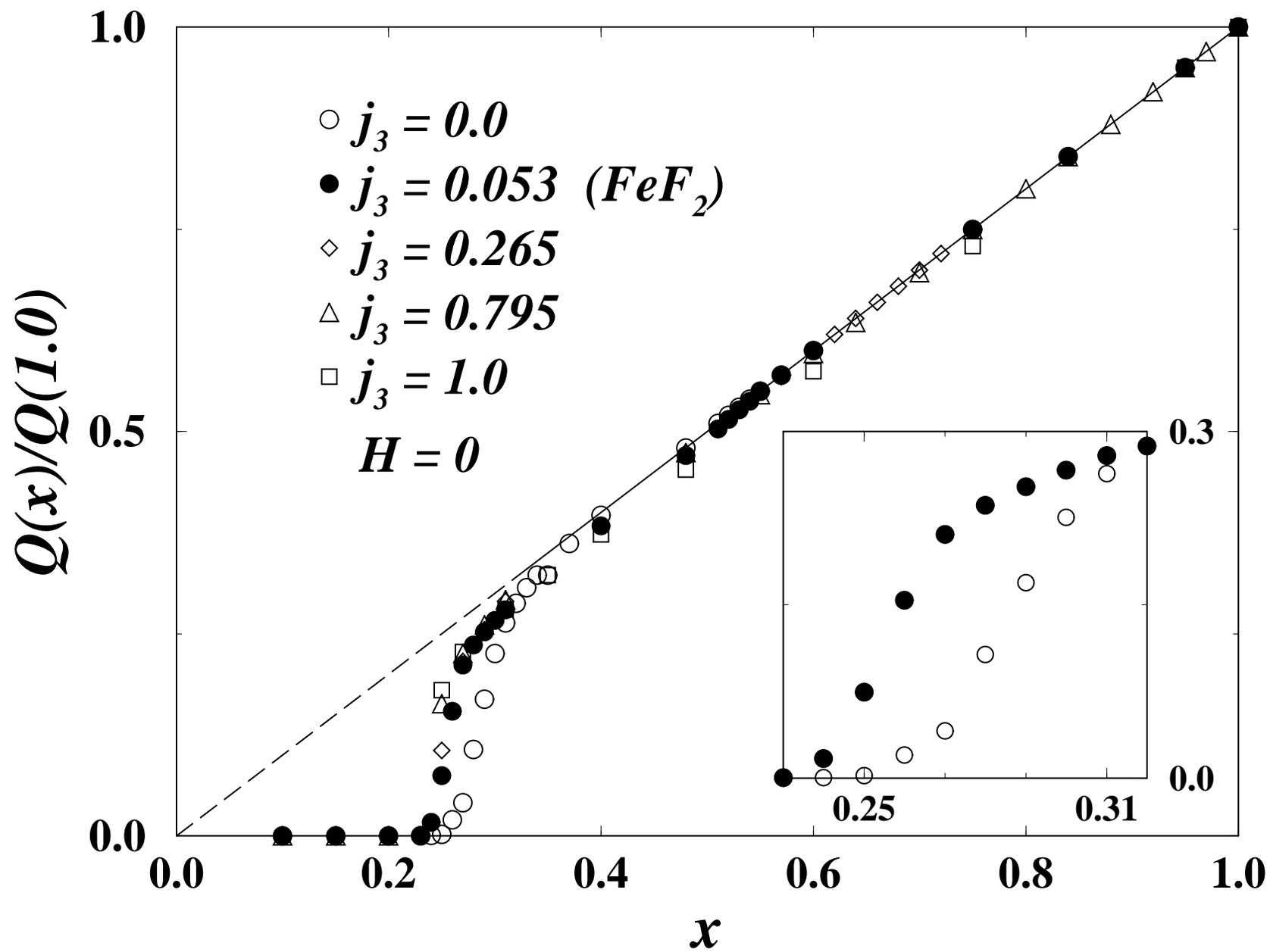
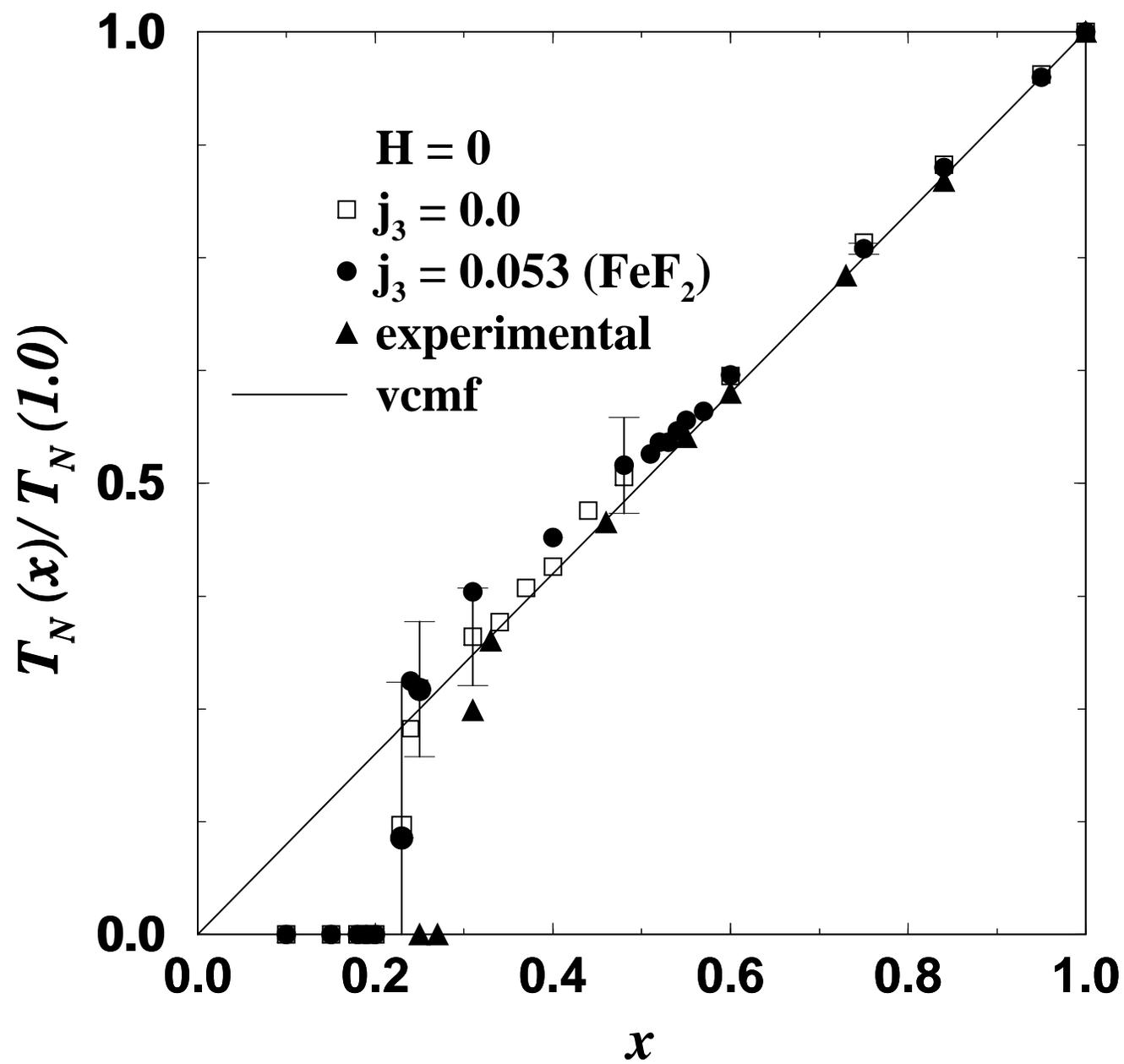
\end{document}